\documentclass[conference,english]{IEEEtran}
\pdfoutput=1					

\usepackage[T1]{fontenc}			
\usepackage[latin9]{inputenc}   	

\usepackage{babel}					

\usepackage{mathrsfs}				

\usepackage{mathtools}				
\usepackage{amsthm}					
\usepackage{amssymb}				

\usepackage{cite}     				
\usepackage{hyperref}				

\usepackage{graphicx} 				

\newtheorem{thm}{Theorem}				
\newtheorem{lem}[thm]{Lemma}			

\renewcommand{\IEEEQED}{\IEEEQEDopen} 	

\DeclareMathOperator{\pr}{Pr} 			
\DeclareMathOperator{\conv}{conv} 		

\begin{document}

\title{On the Power of Cooperation: \\Can a Little Help a Lot?
\\(Extended Version)}

\author{
	\IEEEauthorblockN{
		Parham Noorzad\IEEEauthorrefmark{1},
		Michelle Effros\IEEEauthorrefmark{1},
		Michael Langberg\IEEEauthorrefmark{2},
		Tracey Ho\IEEEauthorrefmark{1}
	}
	\IEEEauthorblockA{
		\IEEEauthorrefmark{1}California Institute of Technology\\ 
		Emails: \{parham, effros, tho\}@caltech.edu
	}
	\IEEEauthorblockA{
		\IEEEauthorrefmark{2}State University of New York at Buffalo\\
		Email: mikel@buffalo.edu
	}
}

\maketitle

\begin{abstract}
In this paper, we propose a new cooperation
model for discrete memoryless multiple access channels.
Unlike in prior cooperation models (e.g., conferencing encoders),
where the transmitters cooperate directly, in this model  
the transmitters cooperate through a larger network. 
We show that under this indirect cooperation model,
there exist channels for which the increase in sum-capacity 
resulting from cooperation
is significantly larger than the rate shared by the transmitters to 
establish the cooperation. This result 
contrasts both with results on the benefit of cooperation 
under prior models and results in the network coding literature, 
where attempts to find examples in which similar 
small network modifications yield large capacity benefits 
have to date been unsuccessful.
\end{abstract}

\section{Introduction}

Cooperation is a potentially powerful strategy in distributed
communication systems. It can both increase the possible transmission rates 
of source messages and improve the reliability of network communications \cite{Kramer}. 
To date, cooperation is not completely understood.
In this paper, we focus on the effect of cooperation
on the capacity region and discuss situations where a small
amount of rate used to enable cooperation results in a large increase in 
the total information that can be carried by the network.

One model of cooperation, proposed by Willems in \cite{Willems}, 
is the \emph{conferencing encoders} (CE) model for the 
discrete memoryless multiple access channel (DM-MAC). In the CE model, 
there is a noiseless link of capacity $C_{12}$ from the first
encoder to the second and a corresponding link of capacity $C_{21}$
back. These links allow a finite number of rounds of communication between the
two encoders;  the total number of bits sent by each encoder
to the other is bounded by the product of the DM-MAC coding blocklength
and the capacity of the encoder's outgoing cooperation link. A similar type of 
cooperation is applied in the broadcast channel with conferencing 
decoders \cite{DaboraServetto} and the interference channel with conferencing 
encoders \cite{MaricEtAl}. More recently, the authors of \cite{PermuterEtAl}
investigate the case where each encoder has partial state information
and conferencing enables information exchange about both the state and 
the messages. 

One can imagine scenarios in which the two transmitters are not able 
to communicate directly or can communicate more effectively through
some other part of the network. The latter can occur, for example, if resources
are less constrained elsewhere in the network than they are for direct
communication. To capture such scenarios, we introduce the 
\emph{cooperation facilitator} (CF) model for the DM-MAC. The cooperation 
facilitator is a node that has complete access to both source messages. 
Based on the messages, it sends limited-rate information to both encoders through a noiseless 
bottleneck link of finite capacity (Figure \ref{fig:networkmodel}). 
We define the \emph{cooperation rate} 
as the capacity of the link carrying the information to be shared. 
One can think of capacity gains obtained 
from this model as an outer bound on the benefit of indirect cooperation.

To study cooperation under this model, we compare the
sum-capacity of a DM-MAC with a CF to the sum-capacity of the DM-MAC
when there is no cooperation between the transmitters. 
This difference equals the 
capacity cost of removing the CF output link from the network.
When the link is removed, the two transmitters are not able to cooperate, and their 
transmitted codewords are independent. We call the resulting network the 
DM-MAC with \emph{independent encoders} (IE). The capacity 
region of this network is due to Ahlswede \cite{Ahlswede1,Ahlswede2}
and Liao \cite{Liao}. 

Since removing the bottleneck link transforms the CF network
into the IE network, the proposed cooperation model is related to 
the edge removal problem in network coding \cite{HoEtAl,JalaliEtAl,LangbergEffros1,LangbergEffros2,LeeEtAl}. 
For networks of noiseless links, there are no known examples of networks for 
which removing a single edge of capacity $\delta$ changes
the capacity region by more than $\delta$ in each dimension, and in some cases it is known
that an impact of more than $\delta$ per dimension is not possible 
\cite{HoEtAl,JalaliEtAl}. Therefore, at least in the situations 
investigated in \cite{HoEtAl,JalaliEtAl}, inserting a cooperation
facilitator in a network cannot increase the sum-capacity
by more than a constant times the cooperation rate.

How much can cooperation help in a DM-MAC? In the CE model, the
increase in sum-capacity is at most the sum of the capacities
of the noiseless links between the two encoders (Section \ref{sec:ce}).
Given the previous discussion, one may wonder whether a
similar result holds for the CF model, that is, whether the
increase in sum-capacity is limited to a constant
times the cooperation rate. In what follows, we see that the
benefit of cooperation can far exceed what might be expected based 
on the CE and edge removal examples. Specifically, we describe
a sequence of DM-MACs with increasing alphabet sizes and set the
cooperation rate for each channel as a function of its alphabet size.
We then show that the increase in sum-capacity that results from cooperation 
grows more quickly than any polynomial function of
the cooperation rate.

In the next section, we review the CE model and its capacity region 
as presented by Willems \cite{Willems}. We give a formal introduction 
to the CF model in Section \ref{sec:cfmodel}.

\section{Prior Work} \label{sec:ce}

Consider the DM-MAC
\begin{equation*}
	\left(\mathcal{X}_{1}\times\mathcal{X}_{2},p_{Y|X_{1},X_{2}}(y|x_{1},x_{2}),\mathcal{Y}\right),
\end{equation*}
where $\mathcal{X}_{1}$, $\mathcal{X}_{2}$, and $\mathcal{Y}$
are finite sets and $p_{Y|X_{1},X_{2}}(y|x_{1},x_{2})$ denotes the
conditional distribution of the output, $Y$, given the inputs, $X_{1}$
and $X_{2}$. 
To simplify notation, we suppress the subscript of the
probability distributions when the corresponding random variables
are clear from context. For example, we write $p(x)$ instead of $p_{X}(x)$. 

There are two sources, source 1 and source 2, whose outputs are the
messages $W_{1}\in\mathcal{W}_{1}=\left\{ 1,\dots,\left\lceil 2^{nR_{1}}\right\rceil \right\} $
and $W_{2}\in\mathcal{W}_{2}=\left\{ 1,\dots,\left\lceil 2^{nR_{2}}\right\rceil \right\} $,
respectively. The random variables $W_{1}$ and $W_{2}$ are independent
and uniformly distributed over their corresponding alphabets. The
real numbers $R_{1}$ and $R_{2}$ are nonnegative and are called
the \emph{message rates}. 

In the IE model, each encoder only has access to its corresponding
message. The encoders are represented by the functions
\begin{align*}
	f_{1n}:\mathcal{W}_{1}&  \rightarrow\mathcal{X}_{1}^{n},\\
	f_{2n}:\mathcal{W}_{2}&  \rightarrow\mathcal{X}_{2}^{n}.
\end{align*}
We denote the output of the encoders by $X_{1}^{n}=f_{1n}(W_{1})$
and $X_{2}^{n}=f_{2n}(W_{2})$. Let $Y^{n}$ be the output of the
channel when the pair $(X_{1}^{n},X_{2}^{n})$ is transmitted.
Using $Y^{n}$, the decoder estimates the original messages via a
decoding function 
$g_{n}:\mathcal{Y}^{n}\rightarrow\mathcal{W}_{1}\times\mathcal{W}_{2}$.

A $\left(2^{nR_{1}},2^{nR_{2}},n\right)$ code for the multiple access
channel is defined as the triple $(f_{1n},f_{2n},g_{n})$.
The average probability of error for this code is given by 
\begin{equation*}
	P_{e}^{(n)}=\pr\left(g_{n}\left(Y^{n}\right)\neq\left(W_{1},W_{2}\right)\right).
\end{equation*}

We say the rate pair $(R_{1},R_{2})$ is achievable if there exists
a sequence of $\left(2^{nR_{1}},2^{nR_{2}},n\right)$ codes such that
$P_{e}^{(n)}$ tends to zero as the blocklength, $n$, approaches infinity.
The capacity region, $\mathscr{C}$, is the closure of the set of all
achievable rate pairs. 

For a given capacity region $\mathscr{C} \subseteq \mathbb{R}_{\geq 0}^{2}$, 
the \emph{sum-capacity} \cite{ElGamalKim},
$C_{\mathrm{S}}$, is defined as 
\begin{equation} \label{eq:sumcapacity}
	C_{\mathrm{S}}=\max\left\{ R_{1}+R_{2}|\left(R_{1},R_{2}\right)\in\mathscr{C}\right\} .
\end{equation}
In the IE model \cite{Ahlswede1, Ahlswede2,Liao}, the sum-capacity is given by
\begin{equation*}
	C_{\mathrm{S-IE}}=\max_{p(x_{1})p(x_{2})}I\left(X_{1},X_{2};Y\right).
\end{equation*}

In the CE model, each encoder shares some information regarding its message 
with the other encoder prior to transmission over
the channel. This sharing of information is achieved through a $K$\emph{-step conference}
over  noiseless links of capacities $C_{12}$ and $C_{21}$. 
A $K$-step conference consists of two sets of functions,
$\left\{ h_{11},\dots,h_{1K}\right\} $ and $\left\{ h_{21},\dots,h_{2K}\right\} $,
which recursively define the random vectors $V_{1}^{K}:=\left( V_{11},\dots,V_{1K}\right)$ 
and $V_{2}^{K}:=\left( V_{21},\dots,V_{2K}\right)$ as
\begin{align*}
	V_{1k} & =h_{1k}\left(W_{1},V_{2}^{k-1}\right),\\
	V_{2k} & =h_{2k}\left(W_{2},V_{1}^{k-1}\right)
\end{align*}
for $k=1,\dots,K$. In step $k$, encoder 1 (encoder 2)
computes $V_{1k}$ ($V_{2k}$) and sends it to encoder 2 (encoder
1). Since the noiseless links between the two encoders are
of capacity $C_{12}$ and $C_{21}$, respectively, we require 
\begin{align*}
	\sum_{k=1}^{K}\log|\mathcal{V}_{1k}| & \leq nC_{12},\\
	\sum_{k=1}^{K}\log|\mathcal{V}_{2k}| & \leq nC_{21}
\end{align*}
where $\mathcal{V}_{ik}$ is the alphabet of the random variable $V_{ik}$
for $i=1,2$ and $k=1,\dots,K$. The outputs of the encoders, $X_{1}^{n}$ and
$X_{2}^{n}$, are given by
\begin{align*}
	X_{1}^{n} & =f_{1n}\left(W_{1},V_{2}^{K}\right),\\
	X_{2}^{n} & =f_{2n}\left(W_{2},V_{1}^{K}\right)
\end{align*}
where $f_{1n}$ and $f_{2n}$ are deterministic functions. 

By studying the capacity region of the CE model \cite{Willems}, we deduce
\begin{equation*}
	C_{\mathrm{S-IE}} \leq C_{\mathrm{S-CE}} \leq C_{\mathrm{S-IE}}+C_{12}+C_{21}.
\end{equation*}
Thus, with conferencing, the sum-capacity increases at most
linearly in $\left(C_{12},C_{21}\right)$ over the sum-capacity
of the IE model. 

\section{The Cooperation Facilitator: Model and Result} \label{sec:cfmodel}
\begin{figure} 
	\begin{center}
		\includegraphics[scale=0.18]{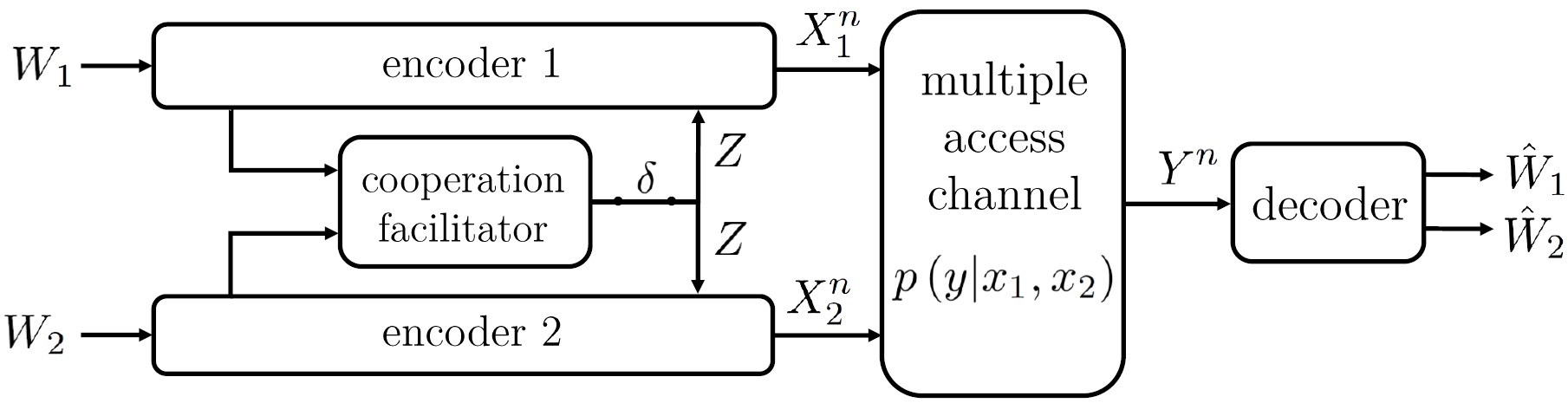}
		\caption{Network model for the DM-MAC with a CF.
		The \emph{cooperation rate} is the capacity of the 
		output link of the CF which we denote by $\delta$.}
		\label{fig:networkmodel}
	\end{center}
\end{figure}

In the CF model, cooperation is made possible not through finite capacity links between the 
encoders, but instead through a third party, the cooperation facilitator, 
which receives information from both encoders and transmits a single 
description of that information back to both (Figure \ref{fig:networkmodel}). 
The cooperation facilitator is represented by the function 
\begin{equation*}
	\phi_{n}:\mathcal{W}_{1}\times\mathcal{W}_{2}\rightarrow\mathcal{Z},
\end{equation*}
where the alphabet $\mathcal{Z}=\left\{ 1,\dots,\left\lceil 2^{n\delta}\right\rceil \right\} $
is determined by the cooperation rate $\delta$. The output of the
cooperation facilitator, $Z=\phi_{n}(W_{1},W_{2})$, is available
to both encoders. 
Each encoder chooses a blocklength-$n$ codeword
as a function of its own source and $Z$ and sends that codeword
to the receiver using $n$ transmissions. Hence the two encoders are
represented by the functions
\begin{align*}
	f_{1n}:\mathcal{W}_{1}\times\mathcal{Z}&  \rightarrow\mathcal{X}_{1}^{n},\\
	f_{2n}:\mathcal{W}_{2}\times\mathcal{Z}&  \rightarrow\mathcal{X}_{2}^{n}.
\end{align*}
The definitions of the decoder, probability of error, and capacity region
are similar to the IE model discussed in the previous section and are omitted. 

Given a pair of functions $f,g:\mathbb{Z}^{+}\rightarrow\mathbb{R}$,
we say $f=o(g)$ if
$\lim_{m\rightarrow\infty}\frac{f(m)}{g(m)}=0.$
We say $f=\omega(g)$ if $g=o(f)$.

For a sequence of DM-MACs 
\begin{equation*}
	\left\{ \left(\mathcal{X}_{1}^{(m)}\times\mathcal{X}_{2}^{(m)},p^{(m)}(y|x_{1},x_{2}),\mathcal{Y}^{(m)}\right)\right\} _{m},
\end{equation*}
let $C_{\mathrm{S-IE}}^{(m)}$ denote the IE sum-capacity of the $m^{\text{th}}$ channel
and  $C_{\mathrm{S-CF}}^{(m)}$ 
denote the CF sum-capacity of the $m^{\text{th}}$ channel when the cooperation rate is $\delta_{m}$.

We are now ready to answer the question  posed in the introduction. In 
the next theorem, which is the main result of this paper, we see that for
a sequence of DM-MACs, the increase in sum-capacity is not only greater 
than the cooperation rate, but also asymptotically larger than any
polynomial function of that rate. In what follows,
$\log(x)$ is the base 2 logarithm of $x$.

\begin{thm} \label{thm:main}
For every sequence of cooperation rates $\{\delta_{m}\}_{m}$ satisfying
$\delta_{m}=\log m +\omega(1)$ and $\delta_{m}\leq m$ and every $\epsilon>0$, 
there exists a sequence of DM-MACs with input alphabets
\begin{equation*}
\mathcal{X}_{1}^{(m)}=\mathcal{X}_{2}^{(m)}=\left\{ 1,\dots,2^{m}\right\},
\end{equation*}
such that for sufficiently large $m$,
\begin{equation*}
	C_{\mathrm{S-CF}}^{(m)}-C_{\mathrm{S-IE}}^{(m)}\geq (3-\sqrt{5+4\epsilon})m-\delta_{m}.
\end{equation*}
For the same sequence of channels, we also have 
\begin{equation*}
	C_{\mathrm{S-CF}}^{(m)}-C_{\mathrm{S-IE}}^{(m)}\leq m+\delta_{m}.
\end{equation*}
\end{thm}

In the above theorem, the choice of $\delta_{m}$ is constrained
only by $\delta_{m}=\log m +\omega(1)$ and $\delta_{m}\leq m$. For example, 
a cooperation rate of $\delta_{m}=\log (m\log m)$ can lead to an 
increase in sum-capacity that is linear in $m$, giving a capacity benefit that
is ``almost'' exponential in the cooperation rate.

\begin{figure}
	\begin{center}
		\includegraphics[scale=0.2]{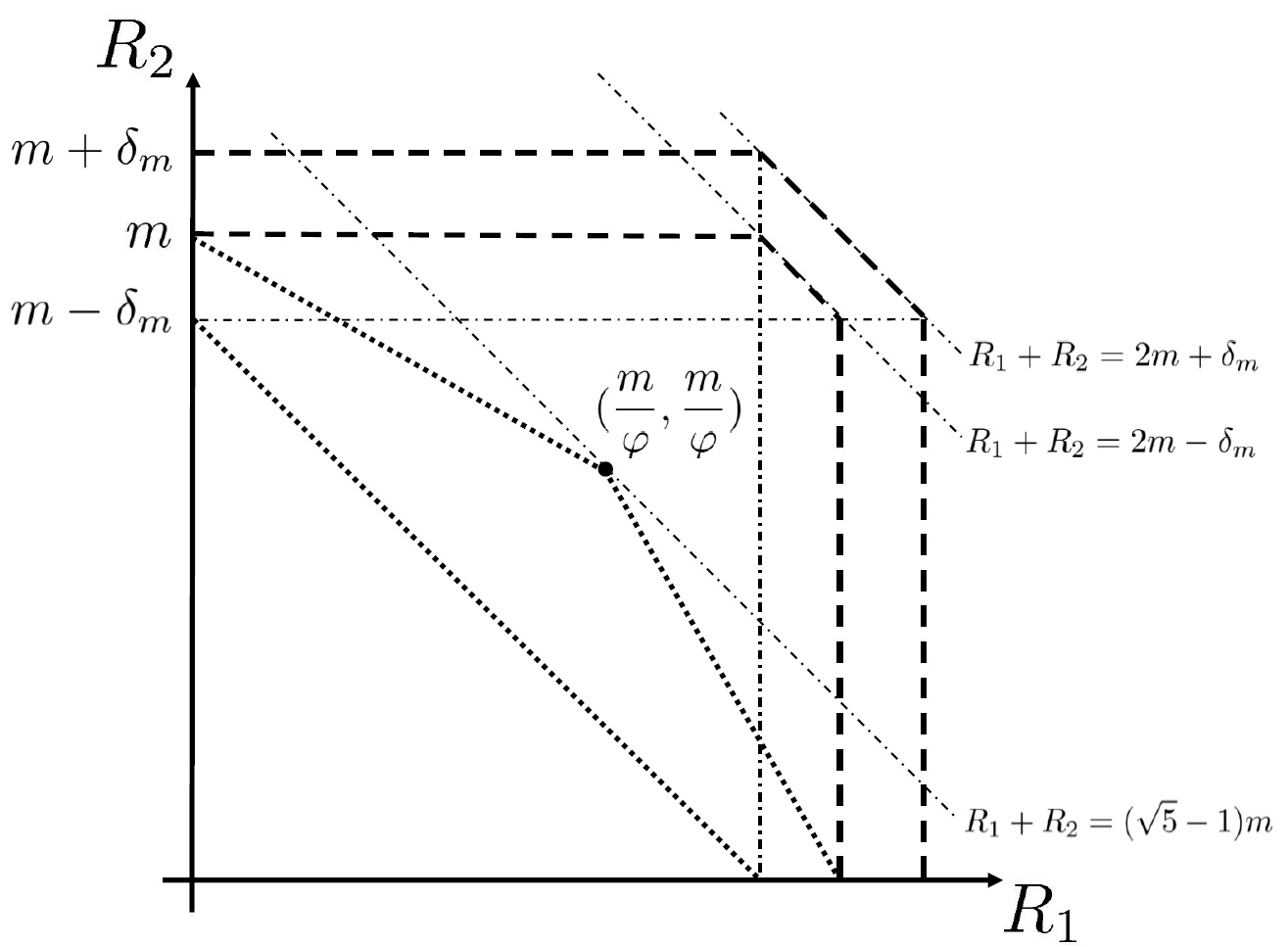}
		\caption
		{
			Inner and outer bounds for the capacity regions of the CF (dashes)
			and IE (dots) models as derived in Sections \ref{sec:CF} and 
			\ref{sec:IE}, respectively. In this figure,
			$\varphi=\frac{1+\sqrt{5}}{2}$.
		}
		\label{fig:capacitybounds}
		\end{center}
\end{figure}

In the next section, we prove the existence of a sequence of DM-MACs 
with properties that are essential for the proof of Theorem \ref{thm:main}.
In Section \ref{sec:CF}, we show that for the sequence 
of channels of Section \ref{sec:chconst},  
\begin{equation} \label{eq:cfbounds}
	2m-\delta_{m} \leq C_{\mathrm{S-CF}}^{(m)}\leq 2m.
\end{equation}
In Section \ref{sec:IE} we show
\begin{equation} \label{eq:iebounds}
	m-\delta_{m}\leq C_{\mathrm{S-IE}}^{(m)}\leq (\sqrt{5+4\epsilon}-1)m.
\end{equation}
Combining these two results gives Theorem \ref{thm:main}. See Figure \ref{fig:capacitybounds}. 
\section{Channel Construction} \label{sec:chconst}

For a fixed positive integer $m$, the channel
\begin{equation*}
	\left( \mathcal{X}_{1}^{(m)}\times \mathcal{X}_{2}^{(m)}, p^{(m)}(y|x_1,x_2), \mathcal{Y}^{(m)} \right)
\end{equation*}
used in the proof of Theorem \ref{thm:main} has input alphabets
$\mathcal{X}_{1}^{(m)}=\mathcal{X}_{2}^{(m)}=\left\{ 1,\dots,2^{m}\right\}$
and output alphabet
\begin{equation*}
\mathcal{Y}^{(m)}=\left( \mathcal{X}_{1}^{(m)}\times\mathcal{X}_{2}^{(m)}\right) \cup\left\{ \left(E,E\right)\right\},
\end{equation*}
where ``$E$'' denotes an erasure symbol. For each 
$(x_1,x_2,y)\in \mathcal{X}_{1}^{(m)}\times\mathcal{X}_{2}^{(m)}\times\mathcal{Y}^{(m)}$,
$p^{(m)}(y|x_1,x_2)$ is defined as a function of the corresponding entry $b_{x_{1}x_{2}}$ 
of a binary matrix $B_{m}=\left(b_{ij}\right)_{i,j=1}^{2^{m}}$.
Precisely,
\begin{equation} \label{eq:chdef}
	p^{(m)}(y|x_{1},x_{2})=\begin{cases}
	1-b_{x_{1}x_{2}}, & \text{if } y=\left(x_{1},x_{2}\right)\\
	b_{x_{1}x_{2}}, & \text{if } y=\left(E,E\right).
\end{cases}
\end{equation}
That is, when $\left(x_{1},x_{2}\right)$ is transmitted, $y=\left(x_{1},x_{2}\right)$ is received
if $b_{x_{1}x_{2}}=0$, and $y=\left(E,E\right)$ is received if $b_{x_{1}x_{2}}=1$.
Thus, we interpret the 0 and 1 entries of $B_{m}$ as ``good'' and ``bad''
entries, respectively. Let $\mathcal{X}^{(m)}=\{1,\dots,2^{m}\}$.
We define the sets
\begin{align*}
	0_{B_{m}} & =\left\{ \left(i,j\right):b_{ij}=0\right\}, \\
	1_{B_{m}} & =\left\{ \left(i,j\right):b_{ij}=1\right\} 
\end{align*}
to be the set of good and bad entries of $\mathcal{X}^{(m)}\times \mathcal{X}^{(m)}$,
respectively. To simplify notation, we drop $m$ as a superscript when it is fixed. 

For every $S,T\subseteq\mathcal{X}$, let $B(S,T)$ be the submatrix
obtained from $B$ by keeping the rows with indices in $S$ and columns
with indices in $T$. For every $x\in\mathcal{X}$, let $B(x,T)=B(\{x\},T)$
and $B(S,x)=B(S,\{x\})$.

The proof of Theorem \ref{thm:main} requires that $B$ satisfies two properties. 
One is that every sufficiently large submatrix of $B$ should have a 
large fraction of bad entries. This property ensures that the sum-capacity
of our channel without cooperation is small (Section \ref{sec:IE}).
The second property is that every submatrix of a specific type should
have at least one good entry. This property enables a significantly higher
sum-capacity under low-rate cooperation using the cooperation facilitator model
(Section \ref{sec:CF}).
Lemma 2 demonstrates that these two properties can be simultaneously achieved. 
A proof of this and all subsequent lemmas can be found in the appendices.
\begin{lem} \label{lem:chconst}
Let $f,g:\mathbb{Z}^{+}\rightarrow\mathbb{Z}^{+}$ be two functions
such that $f(m)=\omega(m)$ and $g(m)=\log m+\omega(1)$. 
Then for every $\epsilon>0$, 
there exists a sequence of $(0,1)$-matrices
$\{B_{m}=\left(b_{ij}\right)_{i,j=1}^{2^{m}}\}_{m}$ such that 

(1) for every $S,T\subseteq{\mathcal{X}^{(m)}}$ that satisfy $|S|,|T|\geq f(m)$,
\begin{equation*}
	\frac{\left|\left(S\times T\right)\cap1_{B_{m}}\right|}{\left|S\right||T|}>1-\epsilon;
\end{equation*}
that is, in every sufficiently large submatrix of $B_{m}$, the fraction
of bad entries is larger than $1-\epsilon$, and

(2) for every $x\in\mathcal{X}^{(m)}$ and $k\in\left\{ 0,1,\dots,2^{m-g(m)}-1\right\} $,
both $B_{m}(x,\mathcal{X}_{m,k})$ and $B_{m}(\mathcal{X}_{m,k},x)$
each contain at least one good entry, where 
\begin{equation*}
	\mathcal{X}_{m,k}=\left\{ k2^{g(m)}+\ell|\ell=1,\dots,2^{g(m)}\right\};
\end{equation*}
that is, if we break each row or column into consecutive blocks 
of size $2^{g(m)}$, each block contains at least one good entry. 
\end{lem}

\textbf{Channel Definition:} Choose functions $f$ and $g$ 
that satisfy the constraints $f(m)=\omega(m)$,
$g(m)=\log m +\omega(1)$, and $\log f(m)=o(m)$. 
Fix a sequence of channels as defined by (\ref{eq:chdef}) 
using matrices $\{B_{m}\}_{m}$ satisfying the properties 
proved possible in Lemma 2 for the chosen functions $f$ and $g$.

\section{Inner and Outer Bounds for the CF Capacity Region}
\label{sec:CF}
For the $m^{\text{th}}$ channel, we show the achievability of the rate pairs
$\left(m,m-g(m)\right)$ and $\left(m-g(m),m\right)$,
with cooperation rate $\delta_{m}=g(m)$. For each, we employ a blocklength-1
code ($n=1$). Time sharing between these codes 
results in an inner bound for the capacity region given by
\begin{align*}
	R_{1},R_{2} & \leq m,\\
	R_{1}+R_{2} & \leq2m-g(m).
\end{align*}

If $R_{1}=m$, $R_{2}=m-g(m)$, and $n=1$, then 
the independent, uniformly distributed messages $W_{1}$ and $W_{2}$
have alphabets $\mathcal{W}_{1}=\{ 1,\dots,2^{m}\} $
and $\mathcal{W}_{2}=\{ 1,\dots,2^{m-g(m)}\} $, respectively.
By the second property of our channel in Lemma 2, for every $w_{1}\in\mathcal{W}_{1}$
and $w_{2}\in\mathcal{W}_{2}$, the submatrix $B_{m}({w_{1},\mathcal{X}_{m,w_{2}-1}})$
contains at least one good entry. Let $z=\phi(w_{1},w_{2})$,
the output of the cooperation facilitator, be an element of 
$\mathcal{Z}=\{ 1,\dots,2^{g(m)}\} $
such that $(w_{1},(w_{2}-1)2^{g(m)}+z)$ is a good entry of 
$B_{m}({w_{1},\mathcal{X}_{m,w_{2}-1}})$. If there's more than one good entry, 
we pick the one that results in the smallest $z$. 

For messages $W_{1}=w_{1}$ and $W_{2}=w_{2}$,
encoder 1 sends $x_{1}=w_{1}$ and encoder 2 sends $x_{2}=(w_{2}-1)2^{g(m)}+z$.
By the definition of our channel (\ref{eq:chdef}), the channel output is
$y=(x_{1},x_{2})$ with probability one, and hence zero
error decoding is possible. Thus the rate pair $(m,m-g(m))$ is achievable.
Note that for this achievability scheme to work, only the second encoder
needs to know the value of $z$. A similar argument proves the achievability
of  $(m-g(m),m)$ and the lower bound of (\ref{eq:cfbounds}) follows.

To find an outer bound for the capacity region, we use the capacity
region of the CE model \cite{Willems} in a special case. Consider the situation 
in which encoder 1 has access to both messages and can transmit information
to encoder 2 on a noiseless link of capacity $\delta_{m}$. Then it is easy
to see that the capacity region of this network contains the capacity
region of the CF model. This situation, however, is equivalent
to the CE model for $C_{12}=\delta_{m}$ and $C_{21}=\infty.$ Hence an outer
bound for the capacity region is given by the set of all rate pairs
$(R_{1},R_{2})$ such that 
\begin{align*}
	R_{1} & \leq I\left(X_{1};Y|X_{2},U\right)+\delta_{m},\\
	R_{1}+R_{2} & \leq I\left(X_{1},X_{2};Y\right)
\end{align*}
for some distribution $p(u)p\left(x_{1}|u\right)p\left(x_{2}|u\right)$.
Note that 
\begin{align*}
	I\left(X_{1};Y|X_{2},U\right)	\leq H\left(X_{1}\right)       & \leq m,\\
	I\left(X_{1},X_{2};Y\right)     \leq H\left(X_{1},X_{2}\right) & \leq 2m,
\end{align*}
and $\delta_{m}=g(m)$, so the region
\begin{align*}
	R_{1} 		& \leq m+g(m),\\
	R_{1}+R_{2} & \leq 2m
\end{align*}
is an outer bound for the CF model. Note that if we switch the 
roles of encoders 1 and 2, we get the outer bound 
\begin{align*}
	R_{2} 		& \leq m+g(m),\\
	R_{1}+R_{2} & \leq 2m.
\end{align*}
Since the intersection of two outer bounds is also an outer bound,
the set of all rate pairs $\left(R_{1},R_{2}\right)$ such that
\begin{align*}
	R_{1},R_{2} & \leq m+g(m),\\
	R_{1}+R_{2} & \leq 2m
\end{align*}
is an outer bound for the CF model as well and the upper bound of
(\ref{eq:cfbounds}) follows. 

\section{Inner and Outer Bounds for the IE Capacity Region}
\label{sec:IE}
Consider the $m^\mathrm{th}$ channel of the construction in Section IV.
In the case where there is no cooperation, we show that the set of
all rate pairs $(R_{1},R_{2})$ satisfying 
\begin{equation*}
	R_{1}+R_{2}\leq m-g(m)
\end{equation*}
is an inner bound for the capacity region. To this end, we
show the achievability of the rate pairs $(m-g(m),0)$ and $(0,m-g(m))$.
The achievability of all other rate pairs in the inner
bound follows by time-sharing between the encoders. Similar to the
achievability result of the previous section, let $n=1$. Then $\mathcal{W}_{1} = \{ 1,\dots,2^{m-g(m)}\} $
and $\mathcal{W}_{2} = \{1\}$.
By our channel construction, for every $w\in\mathcal{W}_{1}$, $B_{m}({\mathcal{X}_{m,w-1},1})$
contains at least one good entry. This means that the first column of $B_{m}$ 
contains at least $|\mathcal{W}_{1}|=2^{m-g(m)}$
good entries. Suppose encoder 1 transmits uniformly on these $2^{m-g(m)}$
good entries and encoder 2 transmits $x_{2}=1$. Then the
input is always on a good entry and the channel output
is the same as the channel input. Thus the pair $(m-g(m),0)$
is achievable. A similar argument shows that the pair $(0,m-g(m))$
is achievable and the inner bound follows. We next find an
outer bound for the IE capacity region.

Let $Y_{1}$ and $Y_{2}$ be the components of $Y$; that is, if $Y=\left(x_{1},x_{2}\right)$,
then $Y_{1}=x_{1}$ and $Y_{2}=x_{2}$, and if $Y=\left(E,E\right)$,
then $Y_{1}=Y_{2}=E$. Note that $Y_{1},Y_{2}\in\mathcal{X}\cup\left\{ E\right\}$.
In the case of independent encoders, $X_{1}$ and $X_{2}$ are independent,
and the distribution of $Y_{1}$ is given by 
\begin{equation} \label{eq:disty1}
	p\left(y_{1}\right)=\begin{cases}
	\gamma_{y_{1}} 	& y_{1}\in\mathcal{X},\\
	1-\gamma 		& y_{1}=E,
	\end{cases}
\end{equation}
where 
\begin{equation*}
	\gamma_{x_{1}}=p\left(x_{1}\right)\sum_{x_{2}:b_{x_{1}x_{2}}=0}p\left(x_{2}\right),
\end{equation*}
for every $x_{1}\in\mathcal{X}$, and $\gamma =\sum_{x_{1}}\gamma_{x_{1}}$.
The capacity region for the IE model (no cooperation) is due to Ahlswede \cite{Ahlswede1,Ahlswede2} and
Liao \cite{Liao}. If $\mathscr{R}_{m}$ is the set
of all pairs $\left(R_{1},R_{2}\right)$ such that 
\begin{align} \label{eq:ieregion}
	R_{1} 		& \leq I\left(X_{1};Y|X_{2}\right),\notag\\
	R_{2} 		& \leq I\left(X_{2};Y|X_{1}\right),\\
	R_{1}+R_{2} & \leq I\left(X_{1},X_{2};Y\right) \notag
\end{align}
for some distribution $p(x_{1})p(x_{2})p(y|x_{1},x_{2})$
and $\conv(A)$ denotes the convex hull of the set $A$, then the
capacity region is given by the closure of $\conv(\mathscr{R}_{m})$.
 
If for all pairs $(R_{1},R_{2})\in \conv(\mathscr{R}_{m})$, 
one of $R_{1}$ or $R_{2}$ is smaller than or equal 
to $\log 2f(m)$,
then the upper bound of (\ref{eq:iebounds})
follows, since 
\begin{equation*}
	R_{1}+R_{2} \leq m+\log 2f(m),
\end{equation*}
and $\log f(m)=o(m)$. On the other hand, if there exist
rate pairs $(R_{1},R_{2})\in \conv(\mathscr{R}_{m})$ such that 
\begin{equation} \label{eq:ratelbound} 
	R_{1},R_{2} > \log 2f(m),
\end{equation}
then by (\ref{eq:ieregion}) and (\ref{eq:ratelbound}),
\begin{equation} \label{eq:entlbound}
	H(X_{1}),H(X_{2}) > \log 2f(m),
\end{equation}
and the following argument shows
\begin{equation*}
	R_{1}+R_{2} \leq (\sqrt{5+4\epsilon}-1)m.
\end{equation*}

For our channel, $Y$, $Y_{1}$, and $Y_{2}$ are
deterministic functions of $(X_{1},X_{2})$,
$(X_{1},Y_{2})$ and $(Y_{1},X_{2})$,
respectively, and the bounds simplify as
\begin{align} \label{eq:entr1}
	R_{1} & \leq I\left(X_{1};Y|X_{2}\right) =H\left(Y_{1}|X_{2}\right)\leq H(Y_{1}), \\
	R_{2} & \leq I\left(X_{2};Y|X_{1}\right) =H\left(Y_{2}|X_{1}\right)\leq H(Y_{2}). \notag
\end{align}

To bound $H(Y_{1})$, we apply the following lemma, proved in 
the appendix. This lemma bounds
the probability that a random variable $X$ falls in a specific set
$T$; the bound is given as a function of the entropy of $X$ and
the cardinality of $T$. For any set $T$, we denote its indicator
function by $\mathbf{1}_{T}$.
\begin{lem} \label{lem:entbound}
Let $X$ be a discrete random variable with distribution $p:\mathcal{X}\rightarrow\mathbb{R}_{\geq 0}$, 
and let $T$ be a subset of $\mathcal{X}$.
If $q:T\rightarrow \mathbb{R}_{\geq 0}$ is a function 
and $\alpha=\sum_{x\in T}q(x)$, then
\begin{equation} \label{eq:entbound1}
	-\sum_{x\in T}q(x)\log q(x)\leq\alpha\log|T|-\alpha\log\alpha. 
\end{equation}
When $q(x)=p(x)\mathbf{1}_{T}(x)$,
the above inequality implies
\begin{equation} \label{eq:entbound2}
	\alpha=\sum_{x\in T}p(x)\leq K\left(1-\frac{H\left(X\right)-1}{\log\left|\mathcal{X}\right|}\right),
\end{equation}
where $K=\left(1-\frac{\log |T|}{\log\left|\mathcal{X}\right|}\right)^{-1}$.
\end{lem}
By (\ref{eq:disty1}),
\begin{equation*}
	H\left(Y_{1}\right) =-\sum_{x_{1}}\gamma_{x_{1}}\log\gamma_{x_{1}}-\left(1-\gamma\right)\log\left(1-\gamma\right).
\end{equation*}
Applying (\ref{eq:entbound1}) from Lemma \ref{lem:entbound},
\begin{equation} \label{eq:enty1}
	H(Y_{1}) \leq\gamma m+H\left(\gamma\right) \leq\gamma m+1.
\end{equation}
We next bound $\gamma$. To this end, we write each of the input distributions 
as a particular convex combination of uniform distributions.
This is stated in the next lemma. 
\begin{lem} \label{lem:convxdist}
If $X$ is a discrete random variable with a finite alphabet $\mathcal{X}$,
then there exists a positive integer $k$, a sequence of positive
numbers $\{ \alpha_{j}\}_{j=1}^{k}$, and a collection of
non-empty subsets of $\mathcal{X}$, $\{ S_{j}\} _{j=1}^{k}$,
such that the following properties are satisfied.

(a) For every $j\in \{1,\dots,k-1\}$, $S_{j+1}$ is a proper subset of $S_j$.
 
(b) For all $x\in\mathcal{X}$, 
\begin{equation*}
p(x)=\sum_{j=1}^{k}\alpha_{j}\frac{\mathbf{1}_{S_{j}}\left(x\right)}{|S_{j}|}.
\end{equation*}

(c) For every $C$, $0<C<|\mathcal{X}|$,
\begin{equation*}
	\sum_{j:|S_{j}|\leq C}\alpha_{j}\leq K\left(1-\frac{H(X)-1}{\log|\mathcal{X}|}\right),
\end{equation*}
where $K=\left(1-\frac{\log C}{\log|\mathcal{X}|}\right)^{-1}$.
\end{lem}
Using the previous lemma we write $p(x_{1})$ and $p(x_{2})$ as
\begin{align*}
	p(x_{1})&=\sum_{i=1}^{k}\alpha_{i}^{(1)}\frac{\mathbf{1}_{S_{i}^{(1)}}\left(x_{1}\right)}{|S_{i}^{(1)}|},\\
	p(x_{2})&=\sum_{j=1}^{l}\alpha_{j}^{(2)}\frac{\mathbf{1}_{S_{j}^{(2)}}\left(x_{2}\right)}{|S_{j}^{(2)}|}.
\end{align*}
Then
\begin{equation*}
	\gamma = \sum_{x_{1},x_{2}:b_{x_{1}x_{2}}=0}p(x_{1})p(x_{2})
		   = \sum_{i=1}^{k}\sum_{j=1}^{l}\alpha_{i}^{(1)}\alpha_{j}^{(2)}\beta_{ij},
\end{equation*}
where 
\begin{align*}
	\beta_{ij} &= \sum_{x_{1},x_{2}:b_{x_{1}x_{2}}=0}\frac{\mathbf{1}_{S_{i}^{(1)}}\left(x_{1}\right)\mathbf{1}_{S_{j}^{(2)}}\left(x_{2}\right)}{|S_{i}^{(1)}||S_{j}^{(2)}|}\\
			   &= \frac{\left|\left(S_{i}^{(1)}\times S_{j}^{(2)}\right)\cap 0_{B_{m}}\right|}{\left|S_{i}^{(1)}\right|\left|S_{j}^{(2)}\right|}
\end{align*}
For every $i$ and $j$, $\beta_{ij}\leq 1$. If, however, $\min\{ |S_{i}^{(1)}|,|S_{j}^{(2)}|\} \geq f(m),$
then by the first property of our channel (Lemma 2), $\beta_{ij}\leq \epsilon$.
Thus by part (c) of Lemma \ref{lem:convxdist} and (\ref{eq:entlbound}),
\begin{align*}
 \gamma &<\epsilon +\sum_{i,j:\min\{ |S_{i}^{(1)}|,|S_{j}^{(2)}|\} < f(m)}\alpha_{i}^{(1)}\alpha_{j}^{(2)}\\
 & = \epsilon +1-\sum_{i,j:\min\{ |S_{i}^{(1)}|,|S_{j}^{(2)}|\} \geq f(m)}\alpha_{i}^{(1)}\alpha_{j}^{(2)}\\
 & = \epsilon +1\\
 & \phantom{=} -\bigg(1-\sum_{i:|S_{i}^{(1)}|< f(m)}\alpha_{i}^{(1)}\bigg)\bigg(1-\sum_{j:|S_{j}^{(2)}|< f(m)}\alpha_{j}^{(2)}\bigg)\\
 & \leq \epsilon +1-\left(1-K_{m}\left(1-\frac{H(X_{1})-1}{m}\right)\right)\\
 & \phantom{\leq} \times\left(1-K_{m}\left(1-\frac{H(X_{2})-1}{m}\right)\right),
\end{align*}
where $K_{m}=\left(1-\frac{\log f(m)}{m}\right)^{-1}$
and $K_{m}\rightarrow 1$
as $m\rightarrow \infty$ since $\log f(m)=o(m)$ by assumption.
Since by (\ref{eq:ieregion}) and (\ref{eq:ratelbound}), 
$\log 2f(m) \leq R_{i}\leq H(X_{i})$ for $i=1,2$,
\begin{align*}
\gamma 	& <\epsilon+1-\left(1-K_{m}\left(1-\frac{R_{1}-1}{m}\right)\right)\\
		& \phantom{=}\times \left(1-K_{m}\left(1-\frac{R_{2}-1}{m}\right)\right)\\
		& =\epsilon+K_{m}\left(2-\frac{R_{1}+R_{2}-2}{m}\right)\\
		& \phantom{=} -K_{m}^{2}\left(1-\frac{R_{1}-1}{m}\right)\left(1-\frac{R_{1}-1}{m}\right).
\end{align*}
Combining the previous inequality with (\ref{eq:entr1}) and (\ref{eq:enty1}) results in 
\begin{align*}
\frac{R_{1}}{m} & \leq\epsilon+\frac{1}{m}+K_{m}\left(2-\frac{R_{1}+R_{2}-2}{m}\right)\\
				& \phantom{=} -K_{m}^{2}\left(1-\frac{R_{1}-1}{m}\right)\left(1-\frac{R_{1}-1}{m}\right)\\
				& =\epsilon+\frac{1}{m}+K_{m}\left(2-\frac{R_{1}+R_{2}-2}{m}\right)\\
				& \phantom{=} -K_{m}^{2}\left(1-\frac{R_{1}+R_{2}-2}{m}+\frac{(R_{1}-1)(R_{2}-1)}{m^{2}}\right).
\end{align*}
If we let $x=\frac{R_{1}}{m}$ and $y=\frac{R_{2}}{m}$, then 
the previous inequality can be written as
\begin{align*}
	x & \leq\epsilon+\frac{1}{m}+K_{m}\left(2+\frac{2}{m}-x-y\right)\\
	  & \phantom{\leq} -K_{m}^{2}\left(1+\frac{2}{m}-x-y+\left(x-\frac{1}{m}\right)\left(y-\frac{1}{m}\right)\right),
\end{align*}
or
\begin{equation} \label{eq:smdef1}
	(x-a_{m})(y+b_{m})\leq c_{m},
\end{equation}
where 
\begin{align*}
	a_{m} & =1+\frac{1}{m}-\frac{1}{K_{m}},\\
	b_{m} & =-1-\frac{1}{m}+\frac{1}{K_{m}}+\frac{1}{K_{m}^{2}},\\
	c_{m} & =-1-\frac{2}{m}-\frac{1}{m^{2}}+\left(2+\frac{2}{m}\right)\frac{1}{K_{m}}\\
		  & \phantom{=}+\left(\epsilon+\frac{1}{m}\right)\frac{1}{K_{m}^{2}}-a_{m}b_{m}.
\end{align*} 
By symmetry, we can also show 
\begin{equation} \label{eq:smdef2}
	(x+b_{m})(y-a_{m})\leq c_{m}.
\end{equation}
Note that 
\begin{align*}
	a &:=\lim_{m\rightarrow\infty}a_{m}=0, \\
	b &:=\lim_{m\rightarrow\infty}b_{m}=1,\\
	c &:=\lim_{m\rightarrow\infty}c_{m}=1+\epsilon.
\end{align*}
Let $S_{m}$ be the set of all nonnegative $x,y$  
that satisfy (\ref{eq:smdef1}) and (\ref{eq:smdef2}) and $\mathscr{S}_{m}$
be the set of all $(mx,my)$ such that $(x,y)\in S_{m}$.
Then by the arguments of this section, every $(R_{1},R_{2})\in\mathscr{R}_{m}$
that satisfies $R_{1},R_{2}>\log 2f(m)$ is in $\mathscr{S}_{m}$. 
As the capacity region is given by the closure of 
$\conv(\mathscr{R}_{m})$, the definition of sum-capacity (\ref{eq:sumcapacity}) implies 
\begin{align*}
\frac{1}{m}C_{\mathrm{S-IE}}^{(m)} & \leq \frac{1}{m}\max_{(R_1,R_2)\in\conv\left(\mathscr{S}_{m}\right)}(R_1+R_2)\\
 & = \max_{(x,y)\in\conv\left(S_{m}\right)}(x+y).
\end{align*}
Thus
\begin{equation} \label{eq:scie}
\limsup_{m\rightarrow\infty}\frac{C_{\mathrm{S-IE}}^{(m)}}{m}\leq\lim_{m\rightarrow\infty}\max_{(x,y)\in\conv\left(S_{m}\right)}(x+y).
\end{equation}
To find the limit on the right hand side, we make use of the following
lemma proved in the appendix.
\begin{lem} \label{lem:sumcapacity}
Suppose $\left\{ a_{m}\right\} _{m=1}^{\infty}$, $\left\{ b_{m}\right\} _{m=1}^{\infty}$
and $\left\{ c_{m}\right\} _{m=1}^{\infty}$ are three sequences of
real numbers such that $\lim_{m\rightarrow\infty}a_{m}=a$, $\lim_{m\rightarrow\infty}b_{m}=b$,
$\lim_{m\rightarrow\infty}c_{m}=c$, where
\begin{equation*}
	b,c,a+b,ab+c>0,
\end{equation*}
and 
\begin{equation*}
	\sqrt{(a+b)^{2}+4c}>b+\frac{c}{b}.
\end{equation*}
For every positive integer $m$, let $S_{m}$ be defined as above.
Then 
\begin{equation*}
	\lim_{m\rightarrow\infty}\max_{(x,y)\in\conv\left(S_{m}\right)}(x+y)  =  a-b+\sqrt{(a+b)^{2}+4c}.
\end{equation*}

\end{lem}
It is easy to see that the sequences above satisfy the assumptions of
Lemma \ref{lem:sumcapacity}. Thus
\begin{equation*}
\limsup_{m\rightarrow\infty}\frac{C_{\mathrm{S-IE}}^{(m)}}{m} \leq\sqrt{5+4\epsilon}-1,
\end{equation*}
Therefore, for all but finitely many $m$, 
\begin{equation*}
	C_{\mathrm{S-IE}}^{(m)}\leq(\sqrt{5+4\epsilon}-1)m.
\end{equation*}
\section{Conclusion}
In this paper, we present a new model for
cooperation and study its benefits in the case of
the encoders of a DM-MAC. Specifically, we present channels
for which the gain in sum-capacity is ``almost'' exponential
in the cooperation rate. The CF model 
can be generalized to other network settings, and its study
is subject to future work. 
\section*{Acknowledgment}
This material is based upon work supported by 
the National Science Foundation under Grant 
Numbers CCF-1321129, CCF-1018741, CCF-1038578, 
and CNS-0905615.
\appendices
\section{Proof of Lemma \ref{lem:chconst}}
We use the probabilistic method \cite{Spencer}. We assign a probability to every
$2^{m}\times2^{m}$ $(0,1)$-matrix and show that the probability
of a matrix having both properties is positive for sufficiently large
$m$; hence, there exists at least one such matrix. Fix $\epsilon>0$,
and let $B_{m}=\left(b_{ij}\right)_{i,j=1}^{2^{m}}$ be a random matrix
with $b_{ij}\overset{\mathrm{iid}}{\sim}\text{Bernoulli}\left(p\right)$,
where $1-\epsilon<p<1$. Let
\begin{equation*}
	\Gamma_{m}=\left\{ S:S\subseteq\mathcal{X}^{(m)} , |S|\geq f(m)\right\}
\end{equation*}
For every $S,T\in\Gamma_{m}$, define the event
\begin{equation*}
	E_{S,T}^{(m)}=\left\{ \frac{\left|(S\times T)\cap1_{B_{m}}\right|}{\left|S\right||T|}\leq1-\epsilon\right\} .
\end{equation*}
It follows
\begin{align*}
\MoveEqLeft \pr\bigg(\bigcup_{S,T\in\Gamma}E_{S,T}^{(m)}\bigg)\\
 & \leq\sum_{S,T\in\Gamma}\pr\left(E_{S,T}^{(m)}\right)\\
 & =\sum_{S,T\in\Gamma}\pr\left(\left|(S\times T)\cap1_{B_{m}}\right|\leq\left(1-\epsilon\right)|S||T|\right)\\
 & =\sum_{S,T\in\Gamma}\sum_{k=0}^{\left\lfloor (1-\epsilon)|S||T|\right\rfloor }{|S||T| \choose k}p^{k}(1-p)^{|S||T|-k}\\
 & =\sum_{i,j=f(m)}^{2^{m}}{2^{m} \choose i}{2^{m} \choose j}\sum_{k=0}^{\left\lfloor (1-\epsilon)ij\right\rfloor }{ij \choose k}p^{k}(1-p)^{ij-k}.
\end{align*}
Suppose $\left\{ X_{l}\right\} _{l=1}^{L}$ is a set of independent random variables
such that for each $l$, $X_{l}\in[a_{l},b_{l}]$ with probability one.
If $S=\sum_{l=1}^{L}X_{i}$,
Hoeffding's inequality \cite{Hoeffding} states that for every $u$ smaller or equal to $\mathbb{E}S$,
we have
\begin{equation*}
\pr\left(S\leq u\right)\leq e^{-\frac{2\left(\mathbb{E}S-u\right)^{2}}{\sum_{l=1}^{L}\left(b_{l}-a_{l}\right)^{2}}}.
\end{equation*}
If $\left\{ X_{l}\right\} _{l=1}^{ij}$ is a set of $ij$ independent Bernoulli($p$)
random variables, then for every $l$, $0\leq X_{l}\leq1$, and
\begin{equation*}
	(1-\epsilon)ij \leq pij = \mathbb{E}\left[\sum_{l=1}^{ij}X_{l}\right].
\end{equation*}
Thus Hoeffding's inequality implies
\begin{align*}
\sum_{k=0}^{\left\lfloor (1-\epsilon)ij\right\rfloor }{ij \choose k}p^{k}(1-p)^{ij-k} & =\pr\left(\sum_{l=1}^{ij}X_{l}\leq (1-\epsilon)ij\right) \\
&\leq e^{-2(p-1+\epsilon)^{2}ij}.
\end{align*}
Since ${2^{m} \choose i}\leq2^{mi}$,
\begin{align*}
\MoveEqLeft \sum_{i,j=f(m)}^{2^{m}}{2^{m} \choose i}{2^{m} \choose j}\sum_{k=0}^{\left\lfloor (1-\epsilon)ij\right\rfloor }{ij \choose k}p^{k}(1-p)^{ij-k}\\
 & \leq\sum_{i,j=f(m)}^{2^{m}}2^{m(i+j)}e^{-2(p-1+\epsilon)^{2}ij}\\
 & =\sum_{i,j=f(m)}^{2^{m}}e^{(i+j)m\ln2-2(p-1+\epsilon)^{2}ij}.
\end{align*}
If we define $h:\mathbb{Z}^{2}\rightarrow\mathbb{R}$ as 
\begin{equation*}
	h(i,j)=(i+j)m\ln2-2(p-1+\epsilon)^{2}ij,
\end{equation*}
then for $j\geq f(m)$, 
\begin{align*}
\MoveEqLeft h(i+1,j)-h(i,j)\\
 & =m\ln2-2(p-1+\epsilon)^{2}j\\
 & \leq m\ln2-2(p-1+\epsilon)^{2}f(m)\\
 & =f(m)\left(\frac{m}{f(m)}\ln2-2(p-1+\epsilon)^{2}\right).
\end{align*}
By assumption, 
\begin{equation*}
\lim_{m\rightarrow\infty}\frac{m}{f(m)}=0,
\end{equation*}
so there exists $M_{1}$ such that for all $m>M_{1}$,
\begin{equation*}
\frac{m}{f(m)}<\frac{2}{\ln2}(p-1+\epsilon)^{2}.
\end{equation*}
Therefore, for $m>M_{1}$ and $y\geq f(m)$, $h$ is decreasing with
respect to $i$. As $h$ is symmetric with respect to $i$ and $j$,
for $m>M_{1}$ and $i\geq f(m)$, we also have $h(i,j+1)-h(i,j)<0$.
Thus $h$ is a decreasing function in $i$ and $j$ for $m>M_{1}$
and $i,j\geq f(m)$. Hence for $m>M_{1}$, 
\begin{align*}
\MoveEqLeft \sum_{i,j=f(m)}^{2^{m}}e^{(i+j)m\ln2-2(p-1+\epsilon)^{2}ij}\\
 & \leq\left(2^{m}-f(m)+1\right)^{2}e^{2mf(m)\ln2-2(p-1+\epsilon)^{2}\left(f(m)\right)^{2}}\\
 & <e^{2m\left(1+f(m)\right)\ln2-2(p-1+\epsilon)^{2}\left(f(m)\right)^{2}}\\
 & =e^{2\left(f(m)\right)^{2}\left(\left(1+\frac{1}{f(m)}\right)\frac{m}{f(m)}\ln2-2(p-1+\epsilon)^{2}\right)}.
\end{align*}
The exponent of the right hand side of the previous inequality,
\begin{equation*}
	2\left(f(m)\right)^{2}\left(\left(1+\frac{1}{f(m)}\right)\frac{m}{f(m)}\ln2-2(p-1+\epsilon)^{2}\right),
\end{equation*}
goes to $-\infty$ as $m$ approaches infinity, so 
\begin{equation*}
	\lim_{m\rightarrow\infty}\pr\bigg(\bigcup_{S,T\in\Gamma}E_{S,T}^{(m)}\bigg)=0.
\end{equation*}
This means that the probability that the fraction of bad entries in a sufficiently
large submatrix is less than $1-\epsilon$ is going to zero. 

Next, we calculate the probability that $B_{m}$ doesn't satisfy the second
property. For every $x\in \mathcal{X}^{(m)}$ and $k\in \{1,\dots,2^{m-g(m)}\}$,
define the event 
\begin{equation*}
	E_{x,k}^{(m)} = \left\{\left(0_{B_{m}(x,\mathcal{X}_{m,k})}\cup0_{B_{m}(\mathcal{X}_{m,k},x)}\right)\cap 0_{B_{m}} =\emptyset\right\}.
\end{equation*}
The probability that for every $x$ and $k$ the sets $B_{m}(x,\mathcal{X}_{m,k})$ and 
$B_{m}(\mathcal{X}_{m,k},x)$ don't have at least one good entry equals
\begin{align*}
\MoveEqLeft \pr\bigg(\bigcup_{x,k}E_{x,k}^{(m)}\bigg)\\
 & \leq\sum_{x\in\mathcal{X}^{(m)}}\sum_{k=1}^{2^{m-g(m)}}\pr\left(0_{B_{m}(x,\mathcal{X}_{m,k})}\cap0_{B_{m}}=\emptyset\right)\\
 & +\sum_{x\in\mathcal{X}^{(m)}}\sum_{k=1}^{2^{m-g(m)}}\pr\left(0_{B_{m}(\mathcal{X}_{m,k},x)}\cap0_{B_{m}}=\emptyset\right)\\
 & =2^{2m-g(m)+1}p^{2^{g(m)}}\\
 & =2^{2^{g(m)}\left(\frac{2m-g(m)+1}{2^{g(m)}}+\log p\right)}.
\end{align*}
Since $m=o(2^{g(m)})$, the exponent of the right hand side of the previous inequality,
\begin{equation*}
2^{g(m)}\left(\frac{2m-g(m)+1}{2^{g(m)}}+\log p\right),
\end{equation*}
goes to $-\infty$ as $m\rightarrow\infty$.
This implies that
\begin{equation*}
\pr\bigg(\bigcup_{x,k}E_{x,k}^{(m)}\bigg)
\end{equation*}
goes to zero as $m\rightarrow\infty$. Thus, by the union bound
the probability that the matrix doesn't satisfy either of these properties
is going to zero. Therefore, for large enough $m$, almost every
$(0,1)$-matrix satisfies both properties, though we only
need one such matrix. 

\section{Proof of Lemma \ref{lem:entbound}}
For the first part, if $\alpha=0$, then $q(x)=0$ 
for every $x\in T$ and both sides equal zero. Otherwise,
\begin{align*}
-\sum_{x\in T}q(x)\log q(x) & =-\alpha\sum_{x\in T}\frac{q(x)}{\alpha}\log \frac{q(x)}{\alpha}-\alpha\log \alpha\\
 & \leq \alpha\log|T|-\alpha\log \alpha,
\end{align*}
since $q(x)/\alpha$ is a probability mass function with entropy $\sum_{x\in T}\frac{q(x)}{\alpha}\log \frac{\alpha}{q(x)}$
and alphabet size $|T|$. 

For the second part, if 
\begin{equation*}
	q(x)=p(x)\mathbf{1}_{T}\left(x\right),
\end{equation*}
then by the previous inequality,
\begin{align*}
	-\sum_{x\in T}p(x)\log p(x) &= -\sum_{x\in T}q(x)\log q(x)\\
								&\leq\alpha\log |T|-\alpha\log \alpha,
\end{align*}
where 
\begin{equation*}
	\alpha = \sum_{x\in T}q(x)= \pr(x\in T).
\end{equation*}
Similarly, replacing $\mathcal{X} \setminus T$ with $T$ results in
\begin{align*}
\MoveEqLeft -\sum_{x\in\mathcal{X}\setminus T}p\left(x\right)\log p\left(x\right)\\
 & \leq\left(1-\alpha\right)\log|\mathcal{X}\setminus T|-\left(1-\alpha\right)\log\left(1-\alpha\right).
\end{align*}
Adding the previous two inequalities gives
\begin{align*}
H\left(X\right) & \leq\alpha\log|T|+\left(1-\alpha\right)\log|\mathcal{X}\setminus T|+H\left(\alpha\right)\\
 & \leq\alpha\log|T|+\left(1-\alpha\right)\log\left|\mathcal{X}\right|+1.
\end{align*}
Therefore,
\begin{equation*}
\frac{H\left(X\right)}{\log\left|\mathcal{X}\right|}\leq1+\frac{1}{\log\left|\mathcal{X}\right|}-\left(1-\frac{\log|T|}{\log\left|\mathcal{X}\right|}\right)\alpha,
\end{equation*}
and
\begin{equation*}
\alpha\leq\frac{1-\frac{H\left(X\right)-1}{\log\left|\mathcal{X}\right|}}{1-\frac{\log|T|}{\log\left|\mathcal{X}\right|}}.
\end{equation*}

\section{Proof of Lemma \ref{lem:convxdist}}
Let $k$ be the cardinality of the range of $p:\mathcal{X}\rightarrow\mathbb{R}$.
Then there exists a sequence $\left\{ x_{j}\right\} _{j=1}^{k}$ such
that 
\begin{equation*}
	\left\{ p(x)|x\in\mathcal{X}\right\} =\left\{ p\left(x_{j}\right)|1\leq j\leq k\right\} ,
\end{equation*}
and
\begin{equation*}
	0<p\left(x_{1}\right)<\dots<p\left(x_{k}\right)\leq1.
\end{equation*}
For $j$, $1\leq j\leq k$, define 
\begin{equation*}
	S_{j}=\left\{ x\in\mathcal{X}|p(x)\geq p(x_{j})\right\} ,
\end{equation*}
and let $S_{k+1}=\emptyset$. Then for $j$, $1\leq j\leq k$, $S_{j+1}\subseteq S_{j}$ 
(part a) and
\begin{equation*}
	S_{j}\setminus S_{j+1}=\left\{ x\in\mathcal{X}|p(x)=p(x_{j})\right\} .
\end{equation*}
Thus, the number of $x\in\mathcal{X}$ such that $p(x)=p(x_{j})$ equals
$|S_{j}\setminus S_{j+1}|$. For $j\in\left\{ 2,\dots,k\right\} $,
define 
\begin{equation*}
	\alpha_{j}=|S_{j}|\left(p(x_{j})-p(x_{j-1})\right),
\end{equation*}
and let $\alpha_{1}=|S_{1}|p(x_{1})$. 

In part (b), the left hand side simplifies as
\begin{align*}
\sum_{j=1}^{k}\alpha_{j}\frac{\mathbf{1}_{S_{j}}(x)}{|S_{j}|} & =\sum_{j=1}^{k}\left(p(x_{j})-p(x_{j-1})\right)\mathbf{1}_{S_{j}}(x)\\
														& =\sum_{j=1}^{k}p(x_{j})\mathbf{1}_{S_{j}\setminus S_{j+1}}(x)\\
														& =p(x).
\end{align*}

In (c), if the set $\left\{ j|1\leq j\leq k,|S_{j}|\leq C\right\} $
is empty, then there's nothing to prove. Otherwise, it's a nonempty
subset of $\left\{ 1,\dots,k\right\} $ and thus has a minimum, which
we call $j^{*}$. Then
\begin{align*}
\sum_{j:|S_{j}|\leq C}\alpha_{j} & =\sum_{j=j^{*}}^{k}\alpha_{j}\\
								 & =\sum_{j=j^{*}}^{k}|S_{j}|\left(p(x_{j})-p(x_{j-1})\right)\\
                                 & =\sum_{j=j^{*}}^{k}|S_{j}\setminus S_{j+1}|p(x_{j})-|S_{j^{*}}|p(x_{j*-1})\\
                                 & =\sum_{x\in S_{j^{*}}}p(x)-|S_{j^{*}}|p(x_{j*-1})\\
                                 & \leq\sum_{x\in S_{j^{*}}}p(x).
\end{align*}
By (\ref{eq:entbound2}) of Lemma \ref{lem:entbound},
\begin{align*}
\sum_{x\in S_{j^{*}}}p(x) & \leq \frac{1}{1-\frac{\log|S_{j^{*}}|}{\log\left|\mathcal{X}\right|}}\left(1-\frac{H\left(X\right)-1}{\log\left|\mathcal{X}\right|}\right)\\
                          & \leq \frac{1}{1-\frac{\log C}{\log\left|\mathcal{X}\right|}}\left(1-\frac{H\left(X\right)-1}{\log\left|\mathcal{X}\right|}\right),
\end{align*}
since $|S_{j^{*}}|\leq C$.
\section{Proof of Lemma \ref{lem:sumcapacity}}

Prior to proving Lemma \ref{lem:sumcapacity}, we state and prove
the following lemma. 
\begin{lem} \label{lem:scaux}
Suppose $a$, $b$, and $c$ are three real numbers such
that $b$, $c$, $a+b$, and $ab+c$ are positive, and 
\begin{equation*}
	\sqrt{(a+b)^{2}+4c}> b+\frac{c}{b}.
\end{equation*}
Let $S$ be the set of all pairs $(x,y)$ such that $x,y\geq0$, and
\begin{equation*}
	\begin{cases}
	(x-a)(y+b) & \leq c,\\
	(x+b)(y-a) & \leq c.
	\end{cases}
\end{equation*}
If $x_{0}$ is the unique positive solution to the equation 
\begin{equation*}
	(x-a)(x+b)=c,
\end{equation*}
then 
\begin{equation*}
	\max_{(x,y)\in\conv(S)}(x+y)=2x_{0}.
\end{equation*}
\end{lem}
\begin{figure}
	\begin{center}
		\includegraphics[scale=0.2]{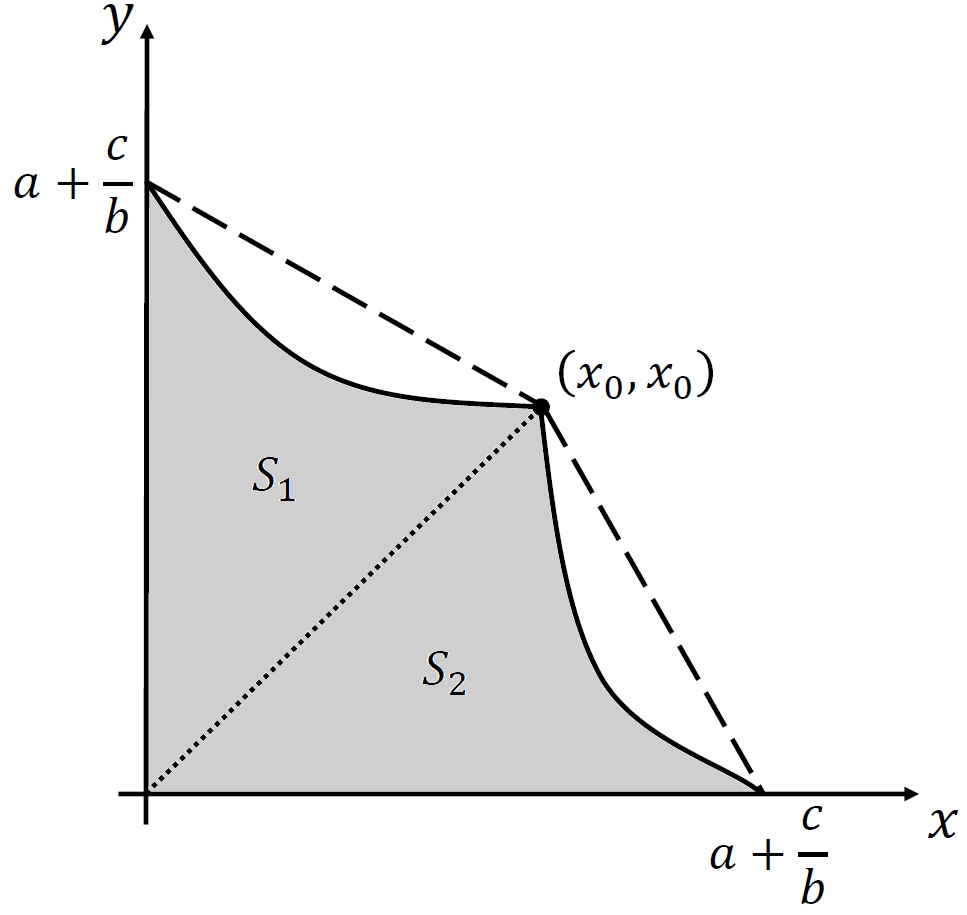}
		\caption{The sets $S_{1}$ and $S_{2}$ (gray area), and their convex hulls.}
		\label{fig:convexhull}
	\end{center}
\end{figure}
\begin{IEEEproof}
Since 
\begin{align*}
	(x-a)(y+b)-(x+b)(y-a) & =(a+b)(x-y)
\end{align*}
and $a+b>0$, the set $S$ can be written as $S=S_{1}\cup S_{2}$ (Figure \ref{fig:convexhull}), 
where $S_{1}$ is the set of all pairs $(x,y)$ such that $0\leq x\leq y$ and 
\begin{equation*}
	(x+b)(y-a)\leq c,
\end{equation*}
and $S_{2}$ is the set of all pairs $(x,y)$ such that $0\leq y\leq x$
and 
\begin{equation*}
	(x-a)(y+b)\leq c.
\end{equation*}
The intersection of $S_{1}$ and $S_{2}$ consists of all pairs $(x,x)$
such that $0\leq x\leq x_{0}$ where 
\begin{equation*}
	x_{0}=\frac{a-b+\sqrt{(a+b)^{2}+4c}}{2}.
\end{equation*}
Note that since $b$, $c$ and $ab+c$ are positive,
\begin{equation*}
	\sqrt{(a+b)^{2}+4c} < a + b + \frac{2c}{b},
\end{equation*} 
so $0<x_{0}<a+\frac{c}{b}$.
The convex hull of $S_{1}$ consists of all pairs $(x,y)$ such that
$0\leq x\leq y$ and
\begin{equation*}
	\left(a+\frac{c}{b}-x_{0}\right)x+x_{0}y\leq\left(a+\frac{c}{b}\right)x_{0},
\end{equation*}
and the convex hull of $S_{2}$ consists of all pairs $(x,y)$ such
that $0\leq y\leq x$ and
\begin{equation*}
	x_{0}x+\left(a+\frac{c}{b}-x_{0}\right)y\leq\left(a+\frac{c}{b}\right)x_{0}.
\end{equation*}
Note that $\conv\left(S_{1}\right)\cup\conv\left(S_{2}\right)$ is the region
bounded by the axes $y=0$, $x=0$, and the piecewise linear function
\begin{equation*}
h(x)=\begin{cases}
\frac{x_{0}-a-\frac{c}{b}}{x_{0}}x+a+\frac{c}{b} & 0\leq x\leq x_{0},\\
\frac{x_{0}}{x_{0}-a-\frac{c}{b}}x-\frac{\left(a+\frac{c}{b}\right)x_{0}}{x_{0}-a-\frac{c}{b}} & x_{0}<x\leq a+\frac{c}{b}.
\end{cases}
\end{equation*}
Since $2x_0\geq a+\frac{c}{b}$ by assumption, 
\begin{equation*}
	\frac{x_{0}-a-\frac{c}{b}}{x_{0}} \geq \frac{x_{0}}{x_{0}-a-\frac{c}{b}}.
\end{equation*}
This means the slope of $h$ is decreasing, or equivalently, $h$ is a concave function.
Thus $\conv\left(S_{1}\right)\cup\conv\left(S_{2}\right)$ is convex. But
\begin{equation*}
	S\subseteq\conv(S_1)\cup\conv(S_2)\subseteq\conv(S),
\end{equation*}
so
\begin{equation*}
	\conv(S)=\conv(S_1)\cup\conv(S_2).
\end{equation*} 
This implies
\begin{equation*}
	\max_{(x,y)\in\conv(S)}(x+y)=2x_{0}.
\end{equation*}
\end{IEEEproof}
Using this lemma, we may prove Lemma \ref{lem:sumcapacity}. There exists a positive $M$ 
such that for every $m\geq M$, 
\begin{equation*}
	b_{m},c_{m},a_{m}+b_{m},a_{m}b_{m}+c_{m}>0
\end{equation*}
and
\begin{equation*}
	\sqrt{(a_{m}+b_{m})^{2}+4c_{m}}-b_{m}-\frac{c_{m}}{b_{m}}>0.
\end{equation*}
 Let $x_{0}^{(m)}$ and $x_{0}$ be the unique positive solutions
to the equations 
\begin{equation*}
	(x_{0}^{(m)}-a_{m})(x_{0}^{(m)}+b_{m})=c_{m}
\end{equation*}
and 
\begin{equation*}
	(x_{0}-a)(x_{0}+b)=c.
\end{equation*}
Since $x_{0}^{(m)}$ and $x_{0}$ are continuous functions of $\left(a_{m},b_{m},c_{m}\right)$
and $(a,b,c)$, respectively, we have
\begin{equation*}
	\lim_{m\rightarrow\infty}x_{0}^{(m)}=x_{0}.
\end{equation*}
Thus by Lemma \ref{lem:scaux},
\begin{align*}
\lim_{m\rightarrow\infty}\max_{(x,y)\in\conv\left(S^{(m)}\right)}(x+y) & =\lim_{m\rightarrow\infty}2x_{0}^{(m)}\\
																	   & =2x_{0}\\
																	   & =a-b+\sqrt{(a+b)^{2}+4c}.
\end{align*}

\bibliographystyle{IEEEtran}
\bibliography{IEEEabrv,ref}{}

\end{document}